\documentclass[notitlepage, 11pt, a4paper]{article}

\usepackage{graphicx}
\usepackage[caption=false]{subfig}
\usepackage{units}

\usepackage{authblk}
\usepackage{amsmath}
\usepackage{textcomp}

\begin{document}


\title{Influence of defect thickness on the angular dependence of coercivity in rare-earth permanent magnets\footnote{This article is published in Applied Physics Letters: 
S. Bance et al., ``Influence of defect thickness on the angular dependence of coercivity in rare-earth permanent magnets'', Appl. Phys. Lett. 104, 182408 (2014); http://dx.doi.org/10.1063/1.4876451}} 



\author{S. Bance}

\author{H. Oezelt}
\author{T. Schrefl}

\affil{Department of Technology, St P\"{o}lten University of Applied Sciences, Matthias Corvinus-Stra\ss{}e 15, A-3100 St P\"{o}lten, Austria}

\author{G. Ciuta}
\author{N. M. Dempsey}
\author{D. Givord}
\affil{Univ. Grenoble Alpes, Institut N\'{E}EL, F-38042 Grenoble, France}
\affil{CNRS, Institut N\'{E}EL, F-38042 Grenoble, France }

\author{M. Winklhofer}
\affil{Department of Earth and Environmental Science, Lud\-wig\--Maxi\-mi\-lians\--University, \\ 80333 Munich, Germany}

\author{G. Hrkac}
\affil{CEMPS, Harrison Building, University of Exeter, Exeter, EX4 4QF, UK}

\author{G. Zimanyi}
\affil{Department of Physics and Astronomy, UC Davis, One Shields Avenue, Davis, California 95616, USA}

\author{O. Gutfleisch}
\affil{Institute for Materials Science, TU Darmstadt, Petersenstra\ss e 23, \\ 64287 Darmstadt, Germany}

\author{T. G. Woodcock}
\affil{IFW Dresden, Helmholtzstra\ss e 20, 01069 Dresden, Germany}

\author{T. Shoji}
\author{M. Yano}
\author{A. Kato}
\author{A. Manabe}
\affil{Toyota Motor Corp., Toyota City, 471-8572, Japan}



\date{09 May 2014}
\maketitle 


\begin{abstract}
The coercive field and angular dependence of the coercive field of single-grain $\mathrm{Nd_{2}Fe_{14}B}$ permanent magnets are computed using finite element micro\-magnetics. 
It is shown that the thickness of surface defects plays a critical role in determining the reversal process. 
For small defect thicknesses reversal is heavily driven by nucleation, whereas with increasing defect thickness domain wall de-pinning becomes more important. 
This change results in an observable shift between two well-known behavioral models.  

A similar trend is observed in experimental measurements of bulk samples, where a Nd-Cu infiltration process has been used to enhance coercivity by modifying the grain boundaries. 
When account is taken of the imperfect grain alignment of real magnets, the single-grain computed results appears to closely match experimental behaviour.
\end{abstract}


The reduced value of the coercive field seen experimentally in rare-earth permanent magnets with respect to the Stoner-Wohlfarth coherent rotation value \cite{brown1945virtues}, 
may be related to the fact that a defect layer with reduced anisotropy exists at the surface of $\mathrm{Nd_{2}Fe_{14}B}$ \cite{kronmuller1987theory, KronmullerDurstSagawa, woodcock2012understanding}. 
This has been examined by Kronm\"{u}ller and collaborators, within the so-called analytical micromagnetic model \cite{kronmuller1987theory, KronmullerDurstSagawa}. 
From comparison between model predictions and experimental data, it was concluded that Stoner-Wohlfarth ``true nucleation'' at defects governs the demagnetization processes \cite{KronmullerDurstSagawa}. 
The analysis of the temperature dependences of the coercive field in the model gives a thickness of the defect region of the order of 1 nm and demagnetizing field values in excess of 1 T. 
Still considering that reversal is a nucleation phenomenon (full reversal develops from a very small initial nucleus), 
Givord \emph{et al}. concluded that various experimental results are not compatible with S-W true nucleation. 
They suggested the occurrence of a propagation/expansion mechanism, reminiscent of domain wall de-pinning, with thermal activation playing an important role \cite{Givord1990experimental, Givord2003thephysics}.
The angular dependence of the coercive field constitutes an important tool to identify the processes governing magnetization reversal in hard magnetic materials. 
Theoretical models for the angular dependence of the coercive field $H_{\mathrm{c}}$ include the Stoner-Wohlfarth (S-W) model 
$H^{\mathrm{SW}}_{\mathrm{c}}(\theta)=H_{\mathrm{A}}(\mathrm{cos}^{2/3}\theta+\mathrm{sin}^{2/3}\theta)^{-3/2}$
, based on coherent rotation, where $H_{\mathrm{A}}=2K_{1}/\mu_{0}M_{\mathrm{s}}$ is the anisotropy field, $K_{1}$ is the uniaxial magneto\-crystalline anisotropy constant and $M_{\mathrm{s}}$ is the saturation magnetization 
and the Kondorsky model 
$H^{\mathrm{K}}_{\mathrm{c}}(\theta)=H_{\mathrm{p}}/\mathrm{\mathrm{cos}}\theta$
, which was originally derived through consideration of pinning mechanisms at internal defect sites, 
where $H_{\mathrm{p}}$ is the field required to depin a domain wall from the defects (it is implicitly assumed here that $H_{\mathrm{p}} << H_{\mathrm{A}} $). 
Various experimental studies considered the angular dependence of the coercive field in $\mathrm{Nd_{2}Fe_{14}B}$ magnets \cite{KronmullerDurstSagawa, kronmuller1987angular, givord1988angular, Elbaz1991, Cebollada1995} 
A difficulty in the experimental analysis is the fact that angular dependence of the coercive field is substantially flattened by the distribution of 
easy-axis orientation in real magnets \cite{givord1988angular, Elbaz1991, Cebollada1995, rieger1999microstructural}. 
In the case where reversal is governed by S-W nucleation, further flattening of the angular dependence of 
the coercive field can be linked to the reduced influence of defects on the nucleation field at large angles \cite{rieger1999microstructural}.

Recent numerical evidence from atomistic calculations for $\mathrm{Nd_{2}Fe_{14}B}$ sintered magnets\cite{hrkac2010role} suggest that this defect thickness varies between 0.4 nm and 1.6 nm. 
It is also possible that the ground boundary phase itself is weakly ferromagnetic and acts as a soft defect at the surface of the $\mathrm{Nd_{2}Fe_{14}B}$ grains.\cite{SepehriAmin20136622}


In this work, the polyhedral grains in $\mathrm{Nd_{2}Fe_{14}B}$ magnets will be approximated by a cube model. 
In such poly\-hedral magnetic grains reversal begins at the edges, usually a corner, where increased de\-mag\-ne\-tizing field causes localized curling of the magne\-ti\-zation. 
The magnetization reversal process will be computed numerically by solving the equation of motion for the magnetization. 
As opposed to linearized micromagnetic models \cite{kronmuller1987angular} both the linear and non-linear nature of the equations will be taken into account. 
Demagnetizing fields will cause a non-uniform magnetization in the remanent state \cite{schabes1988magnetization, schmidts1991size}, thus requiring to redefine the classical term ``nucleation field''.
We will show that two distinct fields can be defined: The critical field that leads to a reversed nucleus at the corner and the critical field at which the nucleus expands. 
This result from dynamic micromagnetic simulations will be confirmed by numerically computing the energy barrier for magnetization switching. 
At the saddle point the magnetization is already reversed near the corner. Thermal activation of a further increase of the external field will cause an expansion of the reversed nucleus. 
The angular dependence of the coercive field will then be compared to experimental data.

In a layered nano\-composite magnet there are three critical fields to consider;
the nucleation field $H_{\mathrm{n}}^{\mathrm{soft}}$ of the soft layer,  
the nucleation field $H_{\mathrm{n}}^{\mathrm{hard}}$ of the hard layer, 
and
the field $H_{\mathrm{p}}$ required to depin a fully developed or partial domain wall from the boundary between the soft and the hard phase. 
At nucleation the magnetization starts to deviate from the $c$ axis and a partial domain wall, within which the magnetization rotates less than $180^{\circ}$, forms. 
If a defect region with severely reduced uni\-axial ani\-sotropy (e.g.\ $K_{1}^{\mathrm{soft}}=0$) is present, $H_{\mathrm{n}}^{\mathrm{soft}}$ and $H_{\mathrm{p}}$ are always lower than $H_{\mathrm{n}}^{\mathrm{hard}}$, 
so the overall coercive field is determined by Eq. (\ref{equation:critical}). 
\begin{equation}
H_{\mathrm{c}}=\max(H_{\mathrm{n}}^{\mathrm{soft}},H_{\mathrm{p}})
\label{equation:critical}
\end{equation} 

Aharoni \cite{aharoni1960reduction} gives analytical expressions for the nucleation field and coercive field for a one-dimensional micromagnetic model, 
an infinite material with a finite slab of certain thickness and $K_{1}=0$. 
In a two-phase magnet with a main hard phase and a soft layer of finite thickness $t$, $H_{\mathrm{n}}^{\mathrm{soft}}$ can be ana\-lytically estimated from Eq. (\ref{equation:Hn}) as long as $t$ is larger than the hard phase domain wall width $\delta_{hard}$.\cite{kronmullerhilzinger, suess2009exchange} 
Generally, decreasing the soft layer thickness increases the nucleation field. 
\begin{equation}
H_{\mathrm{n}}^{\mathrm{soft}}=\frac{2K_{1}^{\mathrm{soft}}}{\mu_{0}M_{\mathrm{s}}^{\mathrm{soft}}}+\frac{2A^{\mathrm{soft}}\pi^{2}}{4t^{2}\mu_{0}M_{\mathrm{s}}^{\mathrm{soft}}}
\label{equation:Hn}
\end{equation} 
where $M_{\mathrm{s}}^{\mathrm{soft}}$ is the saturation magnetization and $A^{\mathrm{soft}}$ is the exchange constant, both of the soft material. 

The pinning field $H_{\mathrm{p}}$ for a domain wall at the interface between a soft and hard layer, after some simplifications, can be calculated from Eq. (\ref{equation:Hp}). \cite{kronmullergoll, suess2006multilayer} 
\begin{equation}
H_{\mathrm{p}}=\frac{1}{4}\times\frac{2(K_{1}^{\mathrm{hard}}-K_{1}^{\mathrm{soft}})}{\mu_{0}M_{\mathrm{s}}^{\mathrm{soft}}}
\label{equation:Hp}
\end{equation}
where $K_{1}^{\mathrm{hard}}$ and $K_{1}^{\mathrm{soft}}$ are the uniaxial anisotropy constants in the hard phase and soft phase respectively. 
Thus, assuming the magnet to behave as an exchange-spring and with $K_{1}^{\mathrm{soft}}=0$ we can account for a 75\% reduction of $H_{\mathrm{p}}$ with respect to the anisotropy field $H_{\mathrm{A}}=2K_{1}^{\mathrm{hard}}/\mu_{0}M_{\mathrm{s}}^{\mathrm{hard}}$, the theoretical coercive field of the hard phase alone. 
Material parameters for $\mathrm{Nd_{2}Fe_{14}B}$ at room temperature were obtained from the literature as $K_{1}=4.8\mathrm{\,MJ/m^{3}}$, $\mu_{0}M_{\mathrm{s}}=1.59\mathrm{\,T}$ and $A=7.6\mathrm{\,pJ/m}$.\cite{groessinger, sagawa1985magnetic} 
For these material parameters, the interfacial de-pinning field is $\mu_{0}H_{\mathrm{p}}=1.90\mathrm{\,T}$.  %
The domain wall width for the hard material is $\delta_{\mathrm{hard}}=\pi\sqrt{A/K_{1}}=3.95\mathrm{\,nm}$. 



\begin{figure}[htp]
\includegraphics[width=1.0\columnwidth]{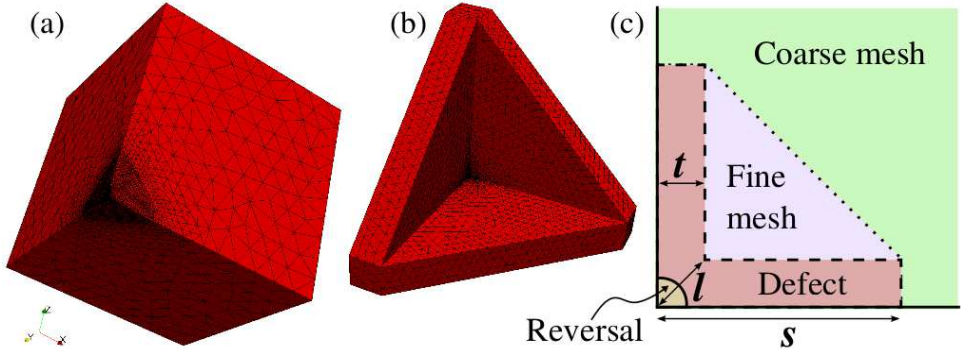}
\caption{(a) The single grain model is geometrically a $L \times L \times L$ cube with a defect shell confined to one corner. (b) The defect shell (seen here from the inside) has thickness $t$ and edge size $s$, measured along the cube edges. (c) A 2D schematic of the defect corner geometry shows the location of increased finite element mesh density, which includes the defect and extends into the hard phase, filling the concave defect cavity. A reversal domain expands from the outer corner of the defect, growing inwards. }
\label{fig:model}
\end{figure}

In this work we consider a single-grain model of cubic geometry with sides $L=100\mathrm{\,nm}$ and a shell-like defect region in one corner only which allows us to locally use a fine mesh for sufficient accuracy (Fig.\ref{fig:model}). 
The defect shell has thickness $t$ and forms a right-angle triangle on each of the three incident cube faces, with adjacent and opposite sides measuring $s=30\mathrm{\,nm}$. 
For Eq. (\ref{equation:Hn}) to be applicable, the domain wall nucleated in the corner of the cube model has to fit into the diagonal space between the outer and inner corners of the defect
so the minimum shell thickness that can fully accommodate a domain wall is $t_{\mathrm{min}}=\sqrt{\delta_{\mathrm{hard}}^{2}/3} = 2.28\mathrm{\,nm}$. 
In the defect region we assume $K_{1}=0$  while $M_{\mathrm{s}}$ and $A$ remain identical to the values in the hard phase. 
The finite element method is used to numerically solve the Landau-Lifschitz-Gilbert (LLG) equation. At each time step we apply a hybrid finite element/boundary element method to compute
the magnetic scalar potential.\cite{schrefl2007numerical} 
A fine mesh is used in the tetrahedral region near the corner so the nucleus and the partial domain wall are located within the fine region before and after depinning. 
Mesh size is constrained by the geometry of the defect, but in all models the maximum element edge size in the fine mesh region is 1.0 nm, below any critical lengths of the materials.\cite{schrefl2007numerical} 
Elsewhere, a coarse mesh size of 10 nm is sufficient. 
The magnet is initially saturated along $+z$ and allowed to relax to its remanent state. 
An external magnetic field $H$ is applied in the opposite direction at a certain field angle $\theta$ from the $-z$ axis in the $x-y$ plane. 
The magneto\-crystalline anisotropy axis is oriented parallel to $z$. 
The field strength is ramped up slowly enough (much slower than the speed of the Larmor procession for the material,\cite{coey2004magnetism}) so that each intermediate state can be considered an equilibrium state. 
$H_{\mathrm{n}}$ is taken to be the field required to flip the polarity of the $z$ -component of the magnetization $M_{z}$ at the very corner of the cube, which corresponds to formation of at least a $90^{\circ}$ domain wall at that corner. 
$H_{\mathrm{c}}$ is measured as the instantaneous external field value at the moment when the centre of the reversal domain wall reaches the edge of the fine mesh region, 
since the rapid expansion of the reversal domain is extremely fast while the field strength changes extremely slowly. 
To find the saddle-point for a specific reversal the nudged elastic band (NEB) method is used to minimize the energy path, taking input from the initial LLG simulations.\cite{schrefl2007numerical} 
A series of NEB simulations can be used to calculate the energy barrier height as a function of $H$, and fitting to Sharrock's law \cite{sharrock1994time} it is possible to estimate $H_{c}$ at a barrier height of $25\mathrm{k}T$, which corresponds to room-temperature stability when an attempt frequency of $f_{0}=10^{10}\mathrm{\,Hz}$ is assumed. 

The angular dependency of coercivity is measured experimentally for four different hot-deformed nano-crystalline NdFeB samples, where a NdCu grain
boundary infiltration process has been applied. A piece of eutectic NdCu alloy is placed on top of the nano-crystalline base material sample and the system is annealed at $600^{\circ}\mathrm{\,C}$. 
This increases the volume fraction of the Nd-rich intergranular boundary phase, increasing coercivity and reducing the remanent magnetization.\cite{SepehriAmin20136622}  
The four samples consist of the base material (no infiltration) and 5\%, 10\% and 20\% infiltration, where these percentages are expressed in terms of outer mass percentage against mass of base materials. 
By infiltration treatment, we have confirmed there is no change in texture degree and grain size. 
The samples have parallelepiped geometry with average dimensions $5\times5\times0.5\mathrm{\,mm}^{3}$. 
After saturation they are rotated about an angle $\theta$ ($0^{\circ}$ to $85^{\circ}$) from the $c$ axis and the field direction is inverted in order to measure the demagnetization curve using a superconducting quantum interference device vibrating sample magnetometer (SQUID-VSM). 
All measurements are made at a temperature of 300 K. 
The coercive field $H_{\mathrm{c}}$ for each angle is taken at the maximum value of 
$dM(\theta)/dH(\theta)$.

\begin{figure}[htp]
\includegraphics[width=0.9\columnwidth]{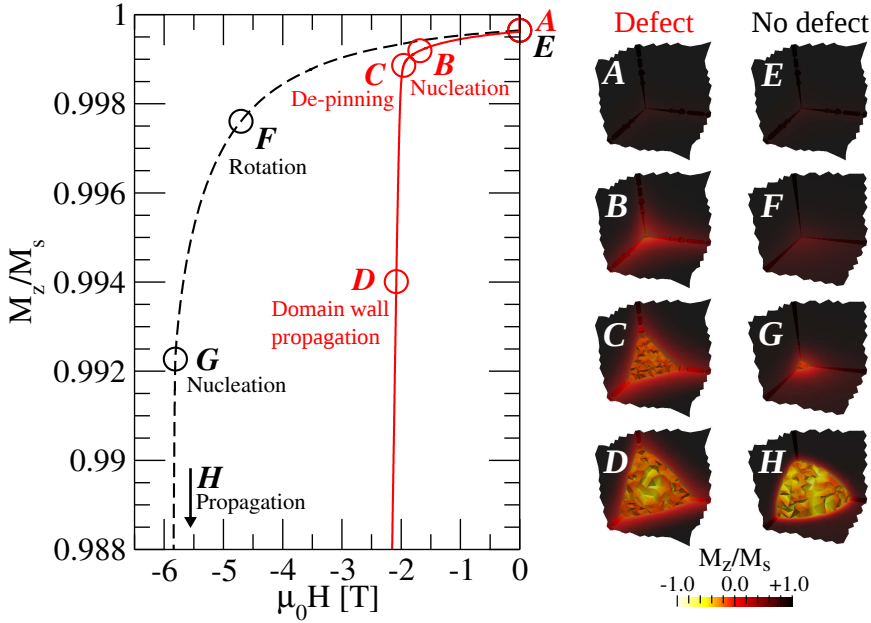}
\caption{Reversal process for both the cube model with a defect shell thickness $t=1.6\mathrm{\,nm}$ and that with no defect shell. $M-H$ reversal curves for both cases show distinct nucleation and pinning mechanisms. }
\label{fig:defect_nodefect}
\end{figure}

Fig.\ref{fig:defect_nodefect} contains a plot and visualizations of the computed reversal for both a grain with shell thickness $t=1.6 \mathrm{\,nm}$ and $t=0.0\mathrm{\,nm}$ (i.e.\ no defect) with $\theta = 0^{\circ}$. 
In both cases the edge inhomogeneities, focussed at the cube corner, are the focal points of initial internal rotation, since the effective field angle is locally reduced. 
This is followed by the nucleation of a reversal domain at the very corner of the cube inside the defect. 
In the defect model a defect-mediated nucleation of a reversal domain inside the shell occurs at an external field strength $\mu _{0}H_{\mathrm{n}}=1.69\mathrm{\,T}$ (B).  
The reversal domain may then expand up to the defect limit, where further expansion is inhibited by an energy barrier.  
At the de-pinning field $\mu _{0}H_{\mathrm{p}}=2.0\mathrm{\,T}$ (C) this energy barrier is overcome, leading to rapid expansion (D) and eventually reversal of the whole grain. 
In the defect-free case, the field required for nucleation is much higher at $5.8\mathrm{\,T}$ (G), so internal rotation continues until a much higher field. 
We note that full reversal of the defect-free cube does not proceed by rotation alone. 

\begin{figure}[htp]
\includegraphics[width=1.0\columnwidth]{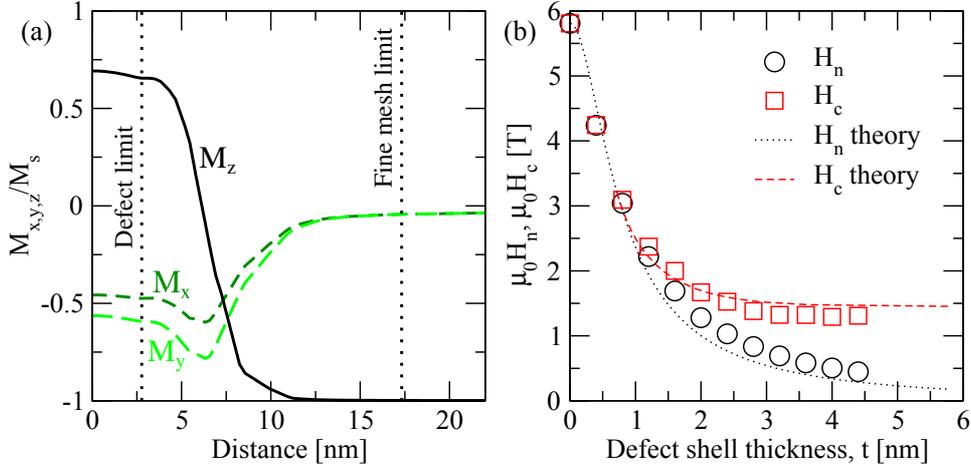}
\caption{(a) Plot of the saddle-point $M_{x}$, $M_{y}$ and $M_{z}$ magnetization components along a diagonal from the reversal corner towards the opposite corner of the cube. The data is taken from the saddle-point configuration of a nudged elastic band (NEB) simulation, which is the highest energy magnetization configuration in the energy-minimized reversal path. (b) A plot of $\mu _{0}H_{\mathrm{n}}$ and $\mu _{0}H_{\mathrm{c}}$ against defect thickness $t$, with Aharoni's 1D defect model for comparison. }
\label{fig:lineHnHc}
\end{figure}

Fig. \ref{fig:lineHnHc}a contains a plot of the saddle-point magnetization components along the inner diagonal from the corner for $t=1.6\mathrm{\,nm}$ and $\theta = 0^{\circ}$.  At the moment of de-pinning the reversal domain already extends past the defect and has entered into the main hard phase. 

Fig. \ref{fig:lineHnHc}b contains a plot of the computed $H_{\mathrm{n}}$ and $H_{\mathrm{c}}$ against $t$. These two field values becomes closer with smaller $t$ until they converge. 
For larger $t$, $H_{\mathrm{c}} > H_{\mathrm{n}}^{\mathrm{soft}}$ so, as per Eq. (\ref{equation:critical}) it is concluded that $H_{\mathrm{c}}=H_{\mathrm{p}}$. 
Likewise, for small $t$ below the convergence value it is concluded that $H_{\mathrm{c}}=H_{\mathrm{n}}$. 
Theoretical plots are included from Aharoni's 1-dimensional nucleation theory for a domain wall in defects.\cite{aharoni1960reduction} 
The equations have been rescaled by $\nicefrac{3}{4}$ to consider the cubic geometry, so that $\mu _{0}H_{\mathrm{a}}=5.81$ T.\cite{newell1998} 
Theoretically, the maximum reduction in $H_{\mathrm{c}}$ in an exchange spring magnet can be achieved with a soft layer of $l=\sqrt{2\pi^{2}A^{\mathrm{soft}}/K_{1}^{\mathrm{hard}}}$, 
corresponding to $l=5.59\mathrm{\,nm}$ and $t=\sqrt{{l^{2}}/3} = 3.23\mathrm{\,nm}$ for our material.\cite{suess2007micromagnetics} 
This is in excellent agreement with Fig. \ref{fig:lineHnHc}b, where above a shell thickness of approximately $t=3\mathrm{\,nm}$ $H_{\mathrm{c}}$ converges to 1.38 T. 

\begin{figure}[htp]
\includegraphics[width=0.8\columnwidth]{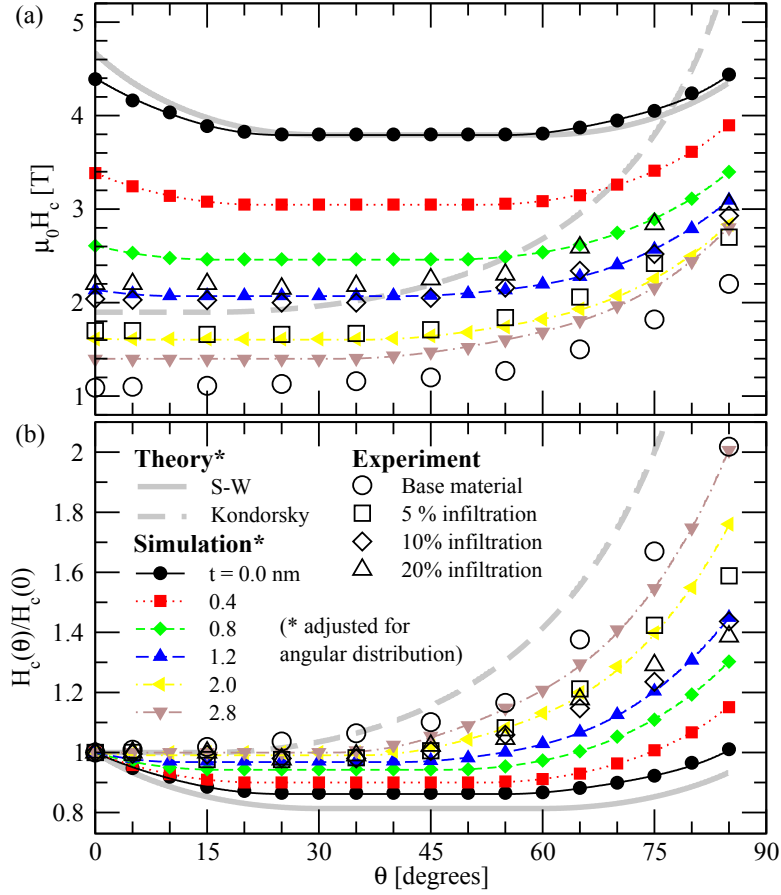}
\caption{(a) Simulation results for angular dependence of coer\-cive field $\mu_{0}H_{\mathrm{c}}$ for $\mathrm{Nd_{2}Fe_{14}B}$ 
at temperature $T=300\mathrm{\,K}$ with varying defect shell thickness $t$ and experimental data for the four samples of varying NdCu infiltration. 
(b) The same data with normalized units. Theoretical plots for the Stoner\--Wohlfarth (S-W) and Kon\-dor\-sky pinning (dashed line) models are given for comparison.  
A simple adjustment to the simulated and theoretical plots has been made to replicate the $15^{\circ}$ angular distribution observed in the experimental samples. }
\label{fig:composite}
\end{figure}

Fig. \ref{fig:composite} contains absolute and normalized plots of the computed $H_{\mathrm{c}}(\theta)$ for various $t$, the experimental data for varying infiltration percentage and the theoretical S-W and Kon\-dor\-sky plots for comparison, 
where for the Kon\-dor\-sky plot we use $H_{\mathrm{c}}(0)=H_{\mathrm{p}}$, from Eq. (\ref{equation:Hp}) and where, 
in order to replicate the effects of grain misalignment, we apply a simple modification rule to the simulated and theoretical data: 
for each field angle $\theta$ the corresponding $H_{\mathrm{c}}$ value is replaced by the lowest value within a certain angular range $\pm \delta$.
A value of $\delta=15^{\circ}$ is chosen to match the angular distribution measured in the experimental samples.\cite{SepehriAmin20136622}  
This adjustment flattens the minimum to a plateau and reduces $H_{\mathrm{c}}$ overall,
giving similar results to the one described in \cite{givord1988angular, Elbaz1991, Cebollada1995} where the angular dependence was calculated for an assembly of imperfectly-oriented exchange-decoupled grains. 
As the value of $t$ increases, a shift to Kon\-dor\-sky\--like behavior is seen, with the minima shifting to lower angles and becoming increasingly less pronounced, 
which is demonstrative of the increased importance of the defect-driven nucleation. 
Above a value of $t=t_{\mathrm{min}} (=2.28\mathrm{\,nm})$, where a full domain wall width is able to fit into the soft layer, the computed plots converge. 
From the experimental data, the base material shows the lowest $H_{\mathrm{c}}$, which becomes larger with increased NdCu in\-fil\-tra\-tion. 
With infiltration treatment, we confirmed no change in both degree of texture and grain size as described before, so that we can control the structure of the grain boundary. 
For the base material a Kon\-dor\-sky-like angular dependence is clearly observed, whereas, with greater infiltration increasing similarity to S-W behavior can be seen, 
with a weak minimum below $45^{\circ}$. 

The progressive shift, from Kondorsky-like to more coherent-rotation like behaviour, may be linked to the associated increase in coercivity \cite{Elbaz1991, Cebollada1995} from non-infiltrated to infiltrated magnets. 
This increase itself can be attributed to two possible pheneomena.  
First, the NdCu infiltration may modify the magnetism of the weakly-ferromagnetic intergrain boundary or second, 
it may reduce surface distortion and weaken the disordered surface defect of reduced anisotropy at the interface between the $\mathrm{Nd_{2}Fe_{14}B}$ grains and the intergrain boundary. 

Experimental results seem to be lower in magnitude than simulation but a direct comparison is not possible since the defect thicknesses of the experimental samples are unknown. 
By increasing further the defect thickness, the coercive field does not decrease much further. 
The presence of a graded interface between the hard phase and the soft defect is also likely to be important. 
Furthermore, thermal activation, which was not considered in the simulations, will reduce co\-er\-ci\-vi\-ty at non-zero temperature.\cite{schrefl2007numerical} 
At 300 K $H_{c}$ was typically reduced by up to 20 \%. 

The results presented here provide compelling evidence that the angle-dependent coercivity behavior of NdFeB magnets is determined by soft defects at the boundaries between the 
main NdFeB grains. 
Using a simple adjustment of the single grain results to replicate the behavior in bulk samples, 
where there are many grains and an angular distribution in their anisotropy angles, we have been able to reproduce the behavior seen experimentally 
and find close agreement with the predictions from simple analytical models of soft defects. 
Recently, a weakly ferromagnetic grain boundary phase in $\mathrm{Nd_{2}Fe_{14}B}$ based magnets was reported.\cite{SepehriAmin20136622} 
One of the possibilities that explains the angular dependence of coer\-civity, which we have simulated by assuming the presence of a surface defect as a mag\-net\-ically soft phase, is that a weakly-ferromagnetic boundary phase may itself act as a surface defect for the neighboring $\mathrm{Nd_{2}Fe_{14}B}$ grain. 
Thus the micro\-magnetic model presented in this paper would still be applicable.
At this moment, the authors are investigating the existence of disordered surface defects and their thicknesses. This should make clear which is the dominant factor for angular dependence of coercivity; the disordered surface defect or the possible weak ferromagnetism of the grain boundary phase.


\section{*Acknowledgments}
We acknowledge the financial support from the Technology Research Association of Magnetic Materials for High Efficient Motors (MagHEM).

\bibliography{BanceAPL}

\end{document}